\documentclass[prx, preprintnumbers, amsmath, amssymb, superscriptaddress, twocolumn]{revtex4-1}
\usepackage{graphicx, color, dcolumn, natbib, bm, amssymb, amsmath,color}
\graphicspath{{fig/}}
\usepackage[final=true]{hyperref}
\usepackage[normalem]{ulem}

\begin{document}
\title{Magnon spectrum of the Weyl semimetal half-Heusler compound GdPtBi}
\author{A. S. Sukhanov}\thanks{alexandr.sukhanov@cpfs.mpg.de}
\affiliation{Max Planck Institute for Chemical Physics of Solids, D-01187 Dresden, Germany}
%\affiliation{Institut f{\"u}r Festk{\"o}rper- und Materialphysik, Technische Universit{\"a}t Dresden, D-01069 Dresden, Germany}
\author{Y. A. Onykiienko}
\affiliation{Institut f{\"u}r Festk{\"o}rper- und Materialphysik, Technische Universit{\"a}t Dresden, D-01069 Dresden, Germany}
\author{R. Bewley}
\affiliation{ISIS Facility, STFC, Rutherford Appleton Laboratory,
Didcot, Oxfordshire OX11-0QX, United Kingdom}
\author{C. Shekhar}
\affiliation{Max Planck Institute for Chemical Physics of Solids, D-01187 Dresden, Germany}
\author{C. Felser}
\affiliation{Max Planck Institute for Chemical Physics of Solids, D-01187 Dresden, Germany}
\author{D. S. Inosov}
\affiliation{Institut f{\"u}r Festk{\"o}rper- und Materialphysik, Technische Universit{\"a}t Dresden, D-01069 Dresden, Germany}
\begin{abstract}

The compound GdPtBi is known as a material where the non-trivial topology of electronic bands interplays with an antiferromagnetic order, which leads to the emergence of many interesting magnetotransport phenomena. Although the magnetic structure of the compound has previously been reliably determined, the magnetic interactions responsible for this type of order remained controversial. In the present study, we employed time-of-flight inelastic neutron scattering to map out the low-temperature spectrum of spin excitations in single-crystalline GdPtBi. The observed spectra reveal two spectrally sharp dispersive spin-wave modes, which reflects the multi-domain state of the $\mathbf{k} = (\frac{1}{2}\frac{1}{2}\frac{1}{2})$ face-centred cubic antiferromagnet in the absence of a symmetry-breaking magnetic field. The magnon dispersion reaches an energy of $\sim 1.1$~meV and features a gap of $\sim 0.15$~meV. Using linear spin-wave theory, we determined the main magnetic microscopic parameters of the compound that provide good agreement between the simulated spectra and the experimental data. We show that GdPtBi is well within the ($\frac{1}{2}\frac{1}{2}\frac{1}{2}$) phase and is dominated by second-neighbor interactions, thus featuring low frustration.

%Despite the half-filled 4$f$ configuration of Gd$3+$ ions, the gap points to a noticeable single-ion anisotropy in GdPtBi, which may be due to Gd 4$f$-5$d$ hybridization.

\end{abstract}

\maketitle

\section{introduction}

Weyl semimetals are a subject of active research owing to their fascinating transport properties such as nonsaturating negative longitudinal magnetoresistance~\cite{Hirsch2016} and magnetothermal resistivity~\cite{Schindler2019}, planar Hall effect~\cite{Kumar2018}, and high carrier mobility~\cite{Shekhar2015}. Their unusual behavior originates in the phenomenon known as the chiral anomaly~\cite{Huang2015}. The electronic band structure of Weyl semimetals allows for the Weyl points, which occur in momentum space in pairs of opposite chiralities. This leads to the emergence of massless-fermion quasiparticle states due to the linear band crossing in the vicinity of the Weyl points~\cite{Lv2015,poprosili}. Promising candidates to Weyl semimetals can be found among Half-Heusler compounds with a general composition XYZ (where X and Y the transition or rare-earth metals, Z the main-group element). Half-Heuslers form a broad family of ternary intermetallic materials that display a vast variety of electronic properties, ranging from semiconductor~\cite{Canfield1991} and semimetallic behaviour~\cite{Kozlova2005} to superconductivity~\cite{Nikitin2015,Pavlosiuk2016} and heavy fermions~\cite{Mun2013,Ueland2014}. Recently, many half-Heusler compounds were predicted to exhibit the topological band inversion, which is a prerequisite for the formation of the Weyl nodes~\cite{Chadov2010}.

Particularly, GdPtBi was recently suggested as a unique compound where Weyl physics coexists with antiferromagnetism (AFM)~\cite{Hirsch2016,Kumar2018,Suzuki2016,Shekhar2018}. Besides the anomalous magnetoresistance, the compound demonstrates a large anomalous Hall effect. The pronounced anomalies in the transverse resistivity occur in a relatively narrow region of applied magnetic fields (with respect to the saturation field) at low temperatures. It was argued~\cite{Suzuki2016,Shekhar2018} that the large anomalous Hall angle can be explained by the specific influences of the AFM order on the electron band structure. However, the interplay between the localized magnetic moments and conduction electrons only partially explained the observed magnetotransport phenomena~\cite{Schindler2019,Suzuki2016}.

In this paper, we address the question on the fundamental interactions between the magnetic Gd$^{3+}$ ions ($S = 7/2$, $L = 0$) in GdPtBi. To construct an appropriate Hamiltonian of the magnetic subsystem, we carried out extensive neutron-spectroscopy measurements. By comparing the experimentally observed magnetic excitations with the results of spin-wave calculations, we quantified the microscopic exchange interactions that characterize the material.

As a representative of the half-Heusler compounds, GdPtBi has a cubic crystal structure with the lattice constant $a = 6.68$~\AA~(space group $F\bar{4}3m$, no. 216) that consists of three interpenetrating fcc lattices~\cite{Canfield1991}. If viewed along the [111] crystallographic direction, the structure is formed by a sequence of the triangular layers of Gd, Pt, and Bi. The AFM order sets in at a N\'{e}el temperature $T_{\text N} = 9$~K. A Curie-Weiss temperature extracted from high-temperature magnetic susceptibility $\theta_{\text{CW}} = -38$~K yields a moderate frustration parameter $f \sim 4$~\cite{Canfield1991,Suzuki2016}. High-field measurements revealed an isotropic saturation field of 25~T at  low temperatures and the saturated magnetic moment of $\sim 6.5\mu_{\text B}$, which is somewhat lower than  the effective paramagnetic moment $\mu_{\text{eff}} = 7.78\mu_{\text B}$~\cite{Suzuki2016,Shekhar2018} and the theoretical value of $7.94\mu_{\text B}$ for the free ion.

\begin{figure}[t]
        \begin{minipage}{0.99\linewidth}
        \center{\includegraphics[width=1\linewidth]{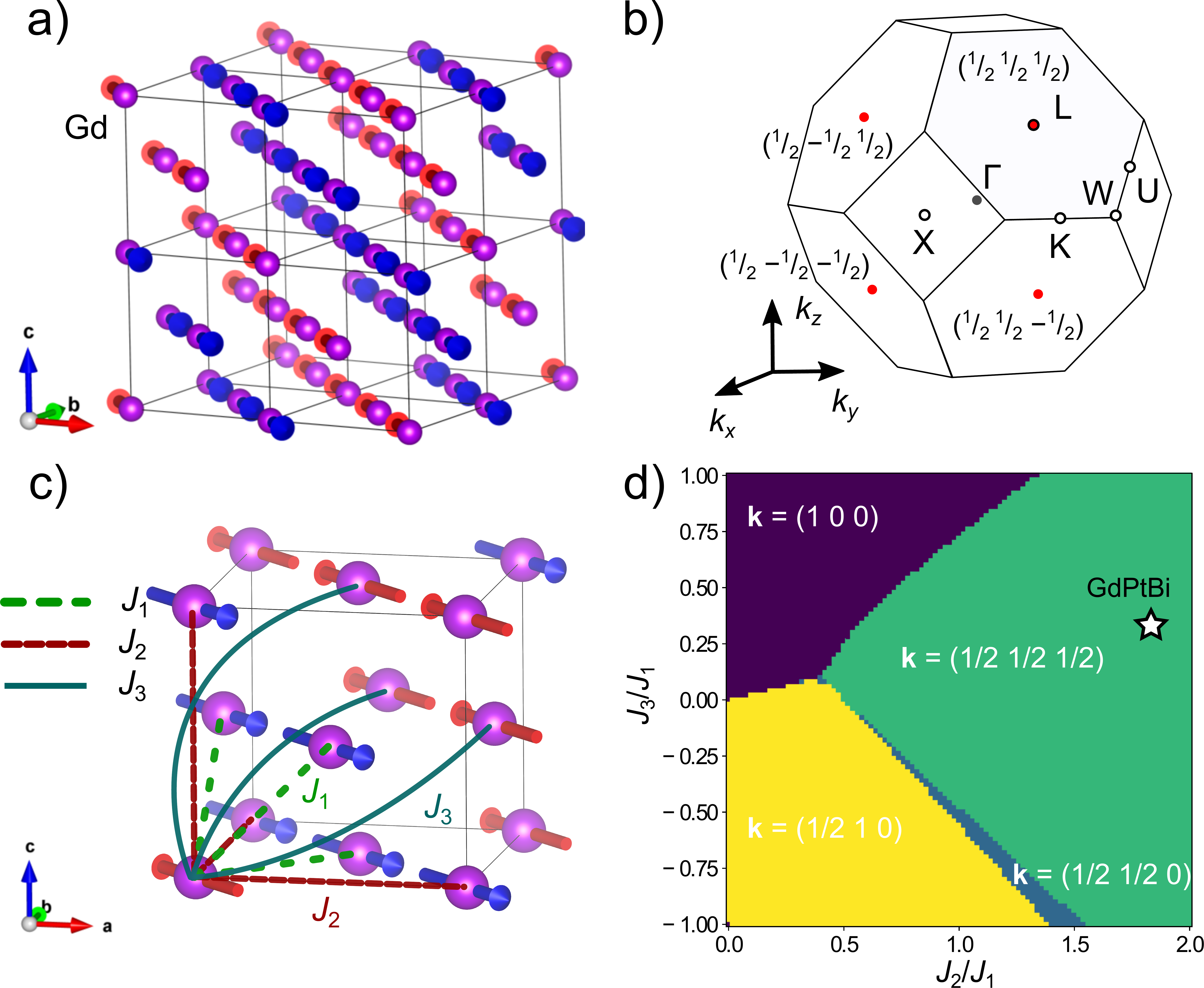}}
        \end{minipage}
        \caption{(color online). (a) The magnetic structure of GdPtBi. Only Gd atoms are shown within the magnetic unit cell, which is eight times larger than the chemical unit cell. (b) The first Brillouin zone of GdPtBi and the propagation vectors that correspond to four AFM domains of the compound. (c) The scheme of the Gd-Gd Heisenberg exchange bonds for up to three coordination spheres. Only Gd atoms are shown within the chemical unit cell. (d) The generic phase diagram of the $J_1$--$J_2$--$J_3$ Heisenberg exchange model of the fcc AFM ($J_1$ is fixed to the AFM sign). The respective propagation vectors label each AFM phase (shaded areas). The position of GdPtBi is marked accordingly to the exchange parameters found in the present study.}
        \label{ris:fig1}
\end{figure}

The magnetic structure of GdPtBi was reported in the previous powder neutron diffraction~\cite{Mueller2014} and single-crystal resonant x-ray scattering~\cite{Kreyssig2011} experiments. It was found that the magnetic moments form ferromagnetic (111) planes stacked antiferromagnetically along the [111] axis (the type-II AFM structure on the fcc lattice). It is also known that the moments are aligned perpendicular to the stacking direction. The magnetic structure is characterized by the propagation vector $\mathbf{k} = (\frac{1}{2}\frac{1}{2}\frac{1}{2})$ and forms a $2\times2\times2$ magnetic supercell, as drawn in Fig.~\ref{ris:fig1}(a). Here and throughout the text, reciprocal-lattice vectors are given in reciprocal lattice units, $1 \text{r.l.u.}=2\pi/a$. The cubic symmetry allows for four magnetic domains with their propagation vectors $\mathbf{k}_i \parallel \langle 111\rangle$. As the magnetic excitations are described in reciprocal space, Fig.~\ref{ris:fig1}(b) illustrates the first Brillouin zone (BZ) of the fcc lattice. The $\mathbf{k}$-vector that corresponds to the order parameter of GdPtBi coincides with the \textit{L}-point on the BZ boundary.

If the nearest-neighbor Heisenberg spins on the fcc lattice have an AFM coupling, four known AFM phases can be stabilized depending on the sign and the strength of the next nearest and the third nearest exchange interactions~\cite{Yildirim1998,Datta2012,Ishizuka2015,Batalov2016}. The exchange interaction scheme for the first three coordination spheres of GdPtBi is depicted in Fig.~\ref{ris:fig1}(c). According to the symmetry, the number of total bonds for the $J_1$, $J_2$, and $J_3$ interaction amounts to 12, 6, and 24, respectively. Figure~\ref{ris:fig1}(d) shows the magnetic phase diagram on the ($J_2/J_1$, $J_3/J_1$) plane. If $J_3$ is absent, the $J_1$--$J_2$ model minimizes the classical energy for the $\mathbf{k} = (100)$ AFM structure for $J_2/J_1 < 0$ ($J_2$ is FM coupling). In this case the magnetic unit cell matches the chemical one. The phase described by $\mathbf{k} = (\frac{1}{2}10)$ is stable when the ratio $J_2/J_1$ satisfies $0 < J_2/J_1 < 1/2$ (for $J_3 = 0$). Otherwise, the $\mathbf{k} = (\frac{1}{2}\frac{1}{2}\frac{1}{2})$ phase is found. If a finite $J_3$ coupling is switched on, the $\mathbf{k} = (100)$ order extends to the positive $J_2/J_1$ ratio. Furthermore, a new phase with the propagation vector $\mathbf{k} = (\frac{1}{2}\frac{1}{2}0)$ appears, which is stable in a much more narrow parameter space. In our work, neutron spectroscopy was applied to determine the position that GdPtBi takes on the phase diagram of the $J_1$--$J_2$--$J_3$ fcc-lattice Heisenberg model.

\section{INS experiment}

Two single crystals of $^{160}$GdPtBi with masses of $\sim 150$ and $\sim 100$ mg were grown using self-flux method as described in Ref.~\cite{Shekhar2018}. Because natural Gd contains a mixture of isotopes with a very large neutron absorption, $^{160}$Gd-isotope enriched (98.5\% enrichment level) pure Gd metal was used for the sample synthesis. To grow the crystals, we followed the exact same procedure that was previously used to synthesize the samples studied in \cite{Schindler2019,Kumar2018,Shekhar2018}. Energy-dispersive x-ray spectra and x-ray Laue diffraction patterns confirmed that the crystals are of the same good quality as the previously studied crystals that contain natural Gd. Inelastic neutron scattering (INS) measurements were conducted at the cold-neutron direct-geometry time-of-flight spectrometer LET~\cite{Bewley2011} located at the ISIS Neutron and Muon Source (Didcot, UK). Using x-ray backscattering Laue, the crystals were oriented in the horizontal $(HHL)$ plane for the measurements and coaligned with the relative misalignment not worse than 0.5$^{\circ}$ to increase the scattering volume. The crystals were mounted onto an aluminium plate holder with a small amount of varnish to minimize the background scattering.

A compromise between the resolution and intensity was achieved by setting the disk chopper frequency to 240~Hz. The incident neutron energy of 11~meV was chosen to cover a sufficiently large part of the 4D momentum-energy reciprocal space. We also used multirep mode~\cite{Russina2009}, which provides two additional datasets with  $E_{\text{i}}$ = 4.80~meV and 2.68~meV to be simultaneously collected. This configuration resulted in an approximate energy resolution at the elastic line $\Delta E$ = 400, 125, and 55~$\mu$eV for the data obtained with the three incident energies, respectively. To map out the reciprocal space, the sample was gradually rotated over 60$^{\circ}$ around the [110] crystallographic direction in 0.5$^{\circ}$ steps. All the measurements were performed at a temperature of 1.6~K, well below $T_{\text{N}}$. The collected data were reduced and analysed using the Horace software~\cite{Ewings2016}.

\section{results}

\begin{figure*}
\includegraphics[width=\linewidth]{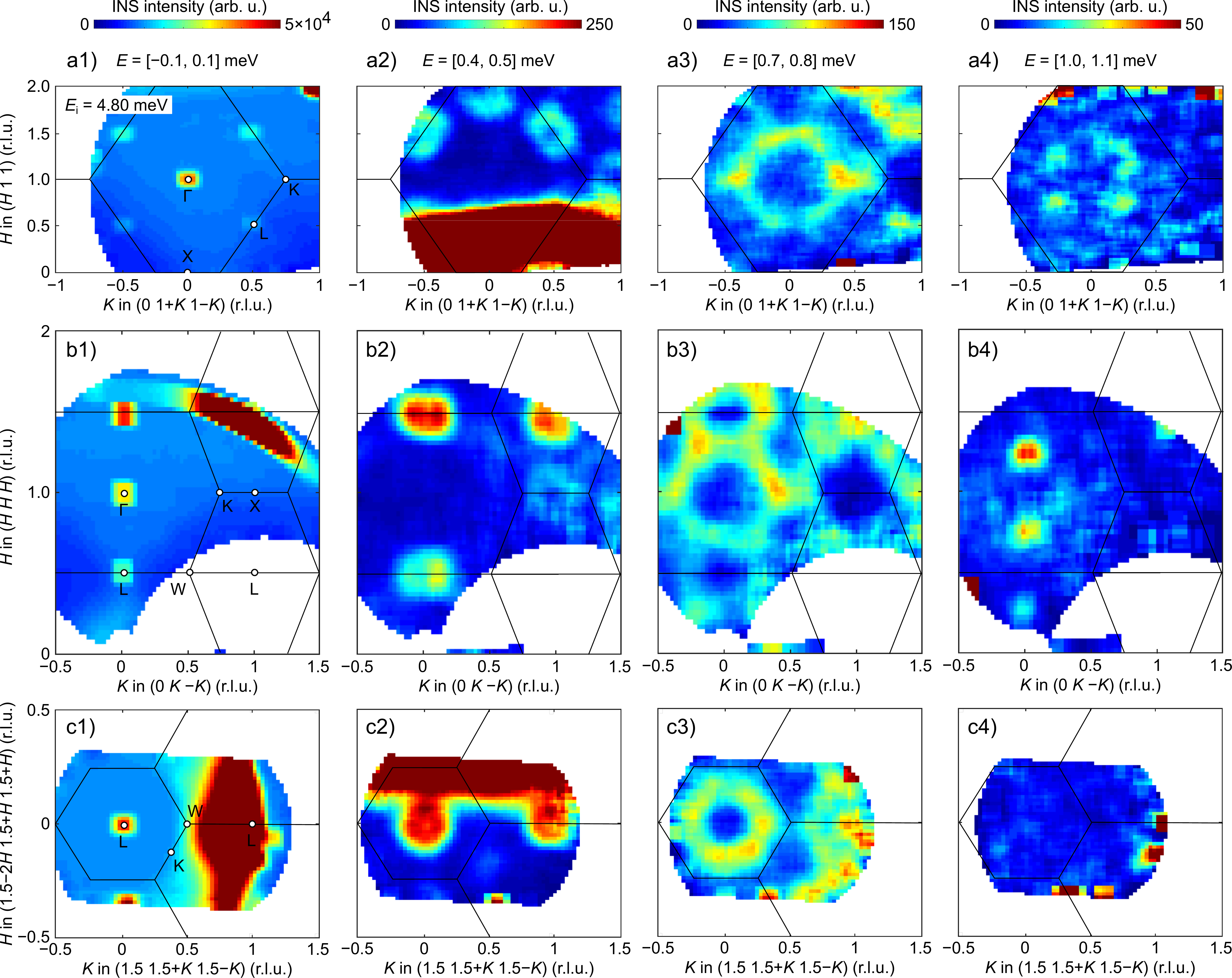}\vspace{3pt}
        \caption{(color online). Constant-energy slices through the time-of-flight INS data collected with the incident neutron energy $E_{\text i} = 4.80$~meV at $T = 1.6$~K. The INS is shown within the ($H$~$1\!+\!K$~$1\!-\!K$) reciprocal-lattice plane (a1)--(a4); ($H$~$H\!+\!K$~$H\!-\!K$) plane (b1)--(b4); ($\frac{3}{2}-\!2H$~$\frac{3}{2}+\!H\!+\!K$~$\frac{3}{2}\!+\!H\!-\!K$) plane (c1)--(c4). The corresponding energy integration intervals are shown on the top. All the momentum slices were integrated over $\pm 0.1$ r.l.u. in the out-of-plane direction. Black solid lines mark the boundaries of the first BZ. Labels in (a1), (b1), and (c1) denote the BZ high-symmetry points. Wide oversaturated areas in (a2), (b1), (c1), and (c2) are measurement artefacts.}
        \label{ris:fig2}
\end{figure*}

\begin{figure}[t]
        \begin{minipage}{0.99\linewidth}
        \center{\includegraphics[width=1\linewidth]{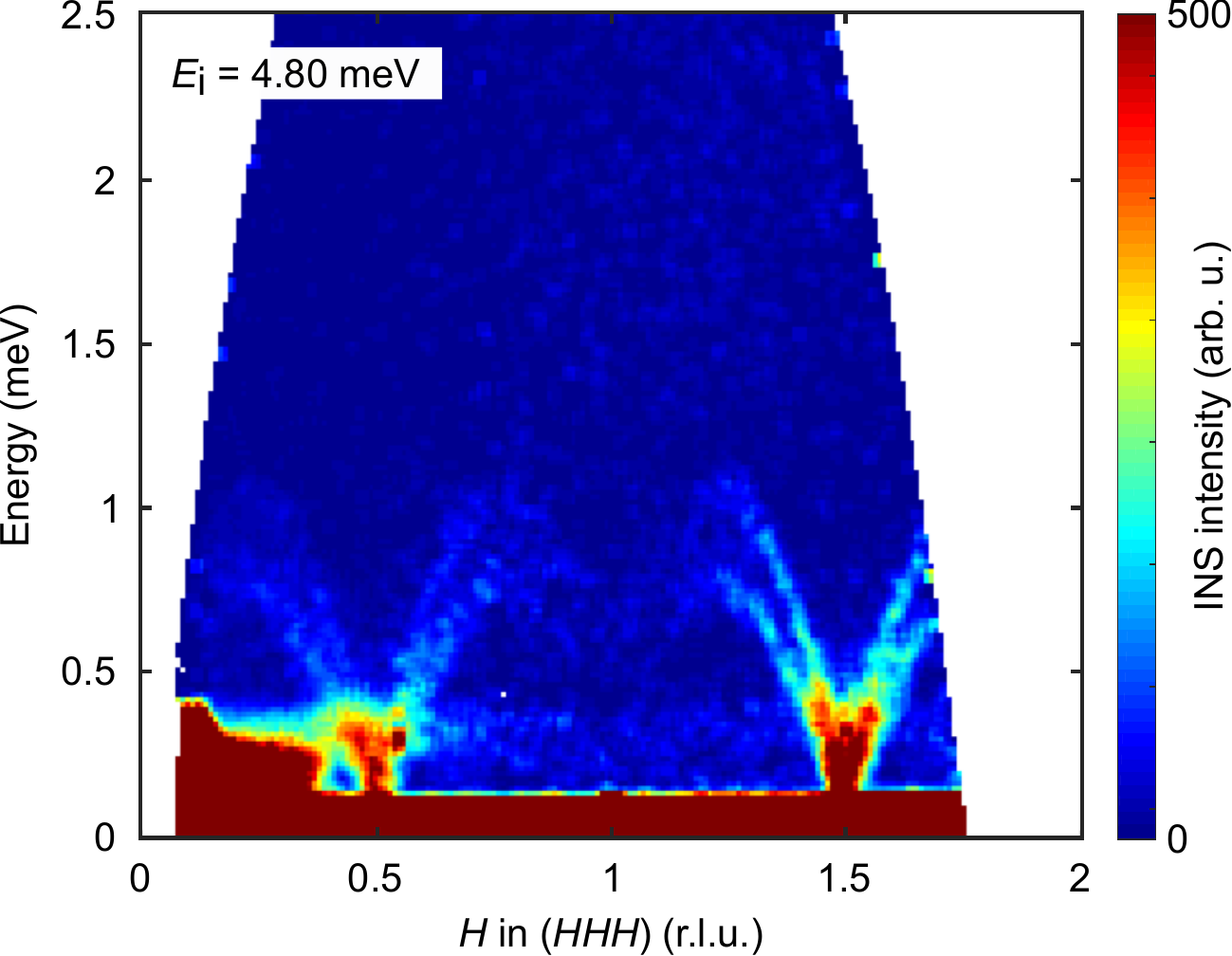}}
        \end{minipage}
        \caption{(color online). The energy-momentum cut through the time-of-flight data collected with the incident neutron energy of 4.80~meV at a temperature of 1.6~K. The magnon dispersion along the ($HHH$) reciprocal direction is shown. The data were integrated over $\pm 0.1$~r.l.u. in two orthogonal directions, which are ($HH-\!2H$) and (0$~\!-\!HH$).}
        \label{ris:fig3}
\end{figure}

\begin{figure}[t]
        \begin{minipage}{0.99\linewidth}
        \center{\includegraphics[width=1\linewidth]{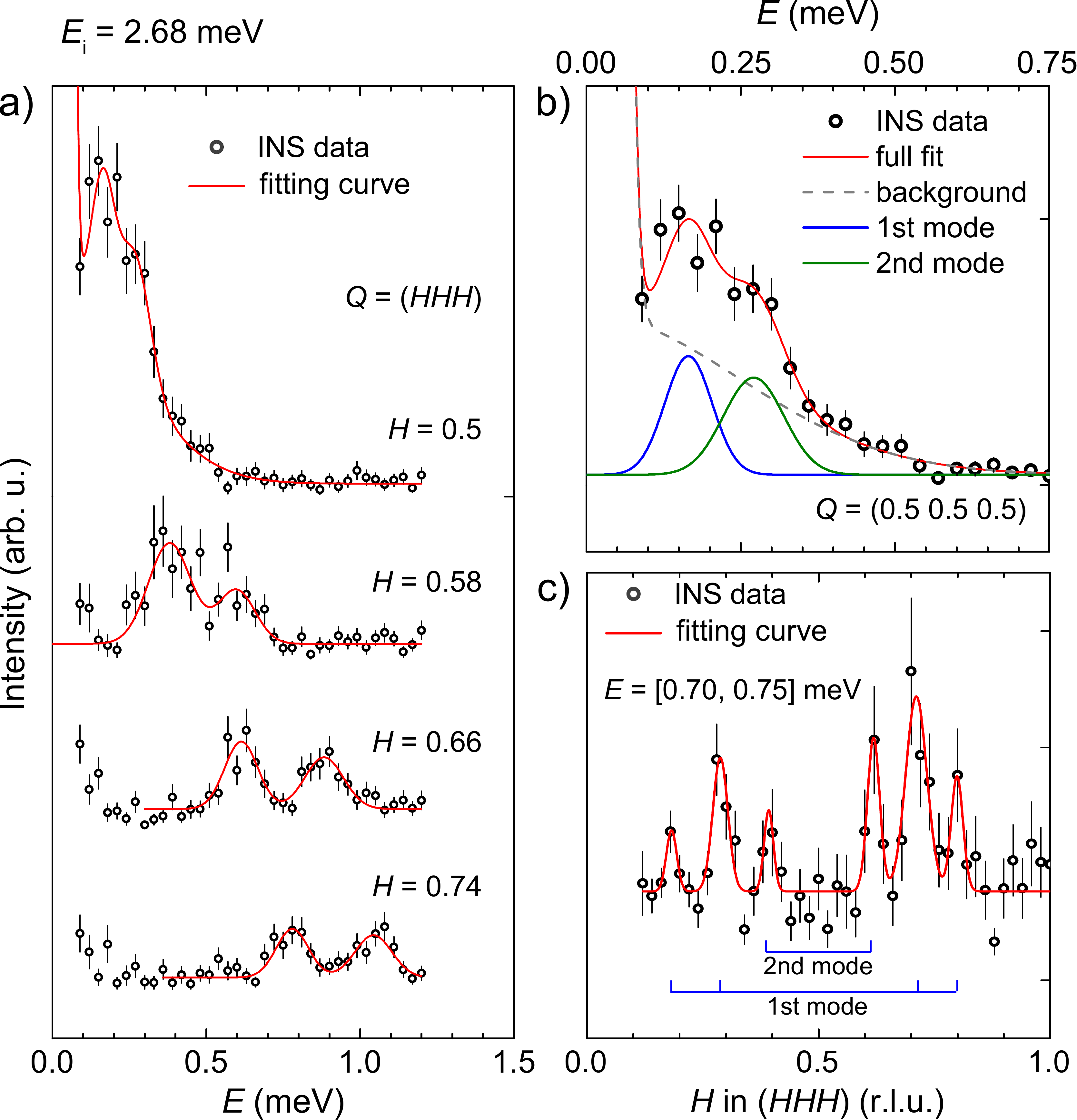}}
        \end{minipage}
        \caption{(color online). (a) Energy profiles of the INS intensity at different momenta along the $(HHH)$ reciprocal direction collected at $E_{\text i} = 2.68$~meV. Solids lines are fits by two Gaussian functions. (b) Details of the profile fit at the $(\frac{1}{2}\frac{1}{2}\frac{1}{2})$ point. Components of the total fit are shown separately for clarity. (c) Momentum profile of the INS intensity for $E = [0.70, 0.75]$~meV. The solid line is a fit by six Gaussian functions.}
        \label{ris:fig4}
\end{figure}

\begin{figure*}
\includegraphics[width=\linewidth]{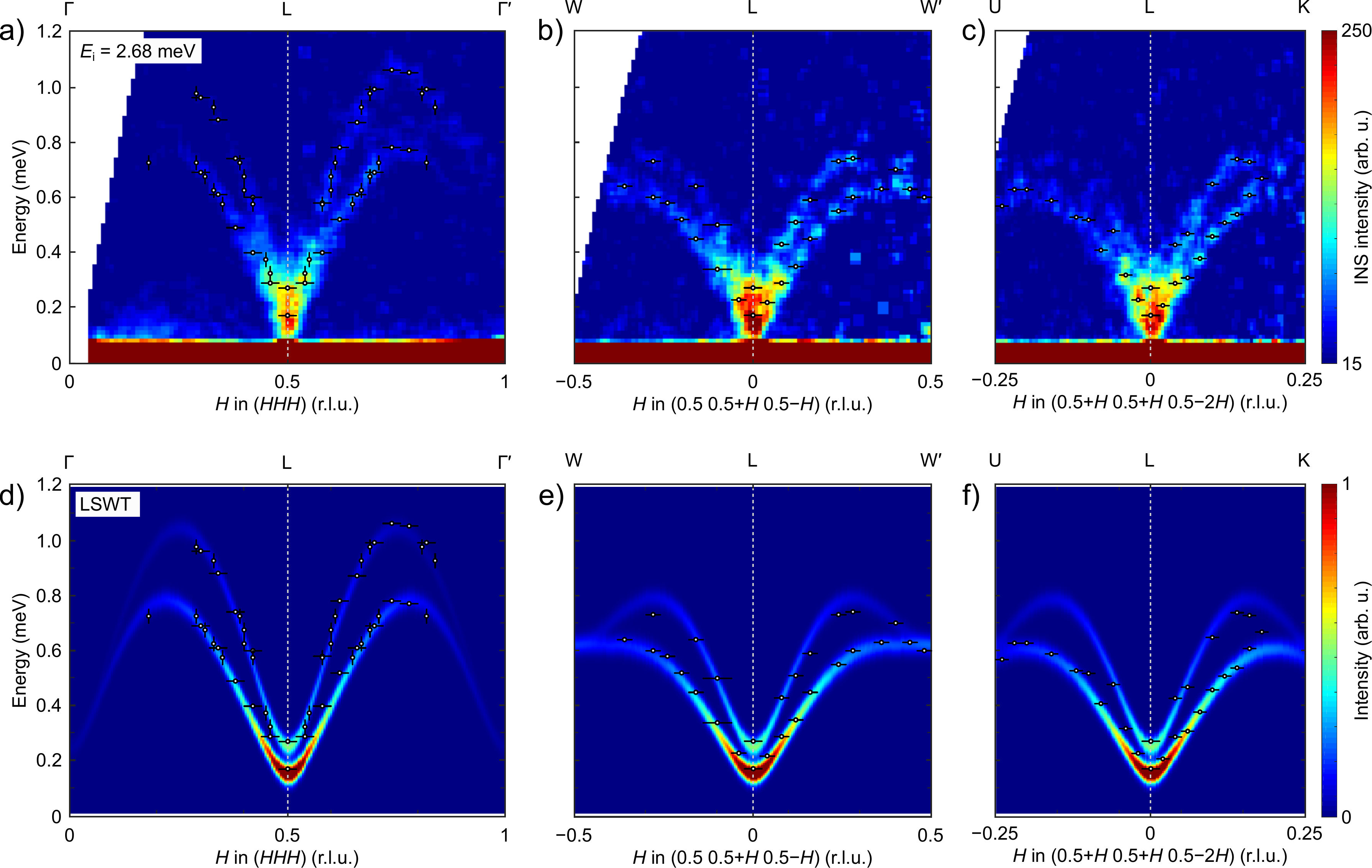}\vspace{3pt}
        \caption{(color online). Magnon spectra for main high-symmetry momentum directions measured with the incident neutron energy of 2.80~meV at $T = 1.6$~K (a)--(c) compared with the results of the LSWT calculations (d)--(f). The data in (a)--(c) were integrated over $\pm 0.1$~r.l.u. in two perpendicular momentum directions. Black circles mark the positions of the magnon dispersion extracted from the Gaussian fits to INS intensity profiles. Horizontal (vertical) intervals denote the momentum (energy) integration range of the fitted constant-$\mathbf{q}$ (constant-$E$) intensity profiles.}
        \label{ris:fig5}
\end{figure*}

First, we turn to the general overview of the observed momentum distribution of the magnon spectral weight at some selected values of the neutron energy transfer. Figures~\ref{ris:fig2}(a1)--\ref{ris:fig2}(c4) summarize the data on the INS intensity extracted from a set of high-symmetry slices in reciprocal space. When the $(H~K~-\!K)$ reciprocal lattice plane is considered [Figs.~\ref{ris:fig2}(a1)--\ref{ris:fig2}(a4)], four of five high-symmetry points of the first BZ (see~Fig.~\ref{ris:fig1}(b)), namely, the zone center $\Gamma$, and the surface-points \textit{K}, \textit{L}, and \textit{X} are found within the plane. The intensity map in Fig.~\ref{ris:fig2}(a1) corresponds to the elastic scattering, this is $E = 0$, where the static structure factor $S(\mathbf{q},0)$ determines the scattering pattern. In accordance with the previously reported magnetic structure, two pairs of magnetic Bragg peaks can be seen at the reciprocal points $(\frac{3}{2}\frac{3}{2}\frac{1}{2})$, $(\frac{1}{2}\frac{1}{2}\frac{3}{2})$, $(\frac{1}{2}\frac{3}{2}\frac{1}{2})$, and $(\frac{3}{2}\frac{1}{2}\frac{3}{2})$, which represent the reflections from the magnetic domains with propagation vectors $\mathbf{k}_1 = (\frac{1}{2}\frac{1}{2}\bar{\frac{1}{2}})$ and $\mathbf{k}_2 = (\frac{1}{2}\bar{\frac{1}{2}}\frac{1}{2})$, respectively. The intense spots at $\Gamma(111)$ and $\Gamma(220)$ are nuclear Bragg peaks of the crystal structure.

After integrating the energy transfer over the range of [0.4, 0.5]~meV [Fig.~\ref{ris:fig2}(a2)], one can observe elliptical features centred at \textit{L} points and circular features located around \textit{X} points. The ellipses of intensity have their major axis oriented along the \textit{K}--\textit{L} reciprocal path, which suggests that the spin-waves are softer in the transverse direction. At a higher energy of [0.7, 0.8]~meV [Fig.~\ref{ris:fig2}(a3)], the cones of intensity stemming from \textit{L} and \textit{X} points expand and merge to form a contour of intensity that resembles the shape of the BZ. The momentum slice of [1.0, 1.1]~meV shows that the whole spectral weight at this energy is located at four isolated points lying in the middle of $\Gamma$--\textit{L} paths [Fig.~\ref{ris:fig2}(a4)]. The constant-energy momentum cuts through the data at $E > 1.1$~meV did not reveal any excitations, which indicates that the top of the magnon band within $(H~K~-\!K)$ plane is located at an energy of 1.1~meV.

Figures~\ref{ris:fig2}(b1)--\ref{ris:fig2}(b4) demonstrate how the momentum-distribution of the magnon spectral weight evolves at the same set of energy slices in another plane in the reciprocal space. This plane is spanned by the orthogonal reciprocal vectors ($01\bar{1}$) and (111) and includes all five high-symmetry BZ points. The elastic cut in Fig.~\ref{ris:fig2}(b1) captures the Bragg peaks from the $\mathbf{k}_3 = (\frac{1}{2}\frac{1}{2}\frac{1}{2})$ domain at $(\frac{3}{2}\frac{3}{2}\frac{3}{2})$ and $(\frac{1}{2}\frac{1}{2}\frac{1}{2})$ points. At $E = [0.4, 0.5]$~meV, two elliptical features around \textit{L}-points at ($HHH$) line are observed. If this pattern is compared with the scattering within ($HK-\!K$) [Fig.~\ref{ris:fig2}(a2)], one can conclude that the dispersion is soft in both the transverse $LK$ and $LW$ directions. Two circular features are observed around the third covered \textit{L}-point at $(\frac{3}{2}\frac{5}{2}\frac{1}{2})$ and the \textit{X}-point at (120). Further, excitations develop in a complex pattern shown in Fig.~\ref{ris:fig2}(b3). This pattern consists of a large hexagonal-shaped feature centred at the $\Gamma$-point, which is connected with two smaller intensity contours that enclose \textit{L}-points. The data collected for the highest magnon energy [Fig.~\ref{ris:fig2}(b4)] shows intense scattering at points $(\frac{5}{4}\frac{5}{4}\frac{5}{4})$, $(\frac{3}{4}\frac{3}{4}\frac{3}{4})$, and $(\frac{1}{4}\frac{1}{4}\frac{1}{4})$.

Figures~\ref{ris:fig2}(c1)--\ref{ris:fig2}(c4) are cuts through a reciprocal-lattice plane that is perpendicular to the ($HHH$) direction and intersects \textit{L}$(\frac{3}{2}\frac{3}{2}\frac{3}{2})$. The selected plane $(\frac{3}{2}-2H$~$\frac{3}{2}+H+K$~$\frac{3}{2}+H-K)$ highlights the spectral weight distribution within the hexagonal face of the BZ, which contains \textit{W}- and \textit{K}-points at the edges and \textit{L}-point in the center. As can be seen in Figs.~\ref{ris:fig2}(c2),~\ref{ris:fig2}(c3), the excitations form quite an isotropic ring of intensity at intermediate energies. The slice at 1.1~meV within this plane [Fig.~\ref{ris:fig2}(c4)] does not reveal any INS intensity. This confirms that the top-energy spin waves are bound within the middle point of $\Gamma L$path, as was observed in cuts shown in Figs.~\ref{ris:fig2}(a4),~\ref{ris:fig2}(b4).

Next, we consider the energy-momentum slice for the momenta along the ($HHH$) reciprocal direction (Fig.~\ref{ris:fig3}). The data collected at $E_{\text i} = 4.80$~meV covers the momenta from $\sim 0.10$ to $\sim 1.75$~r.l.u. at the elastic line and $\sim 0.3$--1.5 r.l.u. at $E = 2.5$~meV. As can be clearly seen, the magnons form two dispersive modes stemming from the magnetic Bragg peaks at $\mathbf{q} = (\frac{1}{2}\frac{1}{2}\frac{1}{2})$ and $(\frac{3}{2}\frac{3}{2}\frac{3}{2})$. Both branches have similar energy at their minima, exhibit a sine-like behavior with approximately 40\% different amplitude, and disperse without intersections. The highest spectral weight is observed in the vicinity of the energy minimum. The INS intensity gradually diminishes towards momenta at which the dispersion acquires the highest energy, which are $\mathbf{q} = (\frac{1}{4}\frac{1}{4}\frac{1}{4})$, $(\frac{3}{4}\frac{3}{4}\frac{3}{4})$ and $(\frac{5}{4}\frac{5}{4}\frac{5}{4})$. A hint to weaker replicas of the same modes but with a vanishing intensity can be noted at $\mathbf{q} < (\frac{1}{4}\frac{1}{4}\frac{1}{4})$. and within $(\frac{3}{4}\frac{3}{4}\frac{3}{4}) < \mathbf{q} < (\frac{5}{4}\frac{5}{4}\frac{5}{4})$. No excitations are observed in the vicinity of the (111) structural reflection. The lower branch reaches an energy of $\sim 0.8$~meV, whereas the upper branch disperse up to $\sim 1.1$~meV. As evidenced by the covered energy range, no other excitations are present at $1.1 < E < 2.5$~(meV). The data collected at $E_{\text i} = 11$~meV revealed absence of any other magnetic excitations up to $\sim 8$~meV. This suggests that the magnetic subsystem of GdPtBi is characterized by spin waves with a bandwidth of $\sim 1$~meV, which is very close to the energy scale given by N\'{e}el temperature of 9~K.

The data collected at $E_{\text i} = 2.68$~meV covers a narrower part of the reciprocal space but provides a better resolution. This enables us to extract the exact energies of the observed spin-wave branches. Figure~\ref{ris:fig4}(a) shows the intensity of INS as a function of energy transfer up to 1.2~meV at a number of fixed momenta along $(HHH)$. As can be seen in the intensity profile at the \textit{L}-point, $(\frac{1}{2}\frac{1}{2}\frac{1}{2})$, the excitations are gapped with a $\sim 0.15$~meV gap. The details on the data analysis are exemplified in Fig.~\ref{ris:fig4}(b), where the result of the profile fit is shown. The measurements show a broad feature (broader than the instrumental energy resolution) in the $\sim 0.10$--0.35~meV range. The profile was fitted by a combination of four Gaussian functions, two of which had their centres fixed at $E = 0$~meV and were used to model the quasielastic background (dotted line in Fig.~\ref{ris:fig4}(b)). The other two Gaussian functions were used to determine the energies of the two magnon modes at the \textit{L}-point (shown as solid lines). The energy gaps of each mode extracted from the fit are $E_1 = 0.17(3)$ and $E_2 = 0.27(6)$~meV. Two branches were resolved and fitted by two Gaussian functions at different momenta as plotted in Fig.~\ref{ris:fig4}(a) for the \textit{L}--$\Gamma$ path. Figure~\ref{ris:fig4}(c) shows an example of the intensity profile as a function of momentum along $(HHH)$ for $0.1 < H < 1$ at the energy integrated within [0.70, 0.75]~meV. Six peaks can be resolved in total, four of which were identified as the first (lower) magnon branch, and the other two peaks are ascribed to the second (upper) mode. 

\begin{table}[b]
\small\addtolength{\tabcolsep}{+7pt}
\caption{Parameters of the model in Eq.~\ref{eq:eq1} providing the best reproduction of the observed magnon spectra (in meV), S = 7/2.}
\label{tab:tab1}
 \begin{tabular}{c c c c}
 \hline\hline
 $J_1S $ & $J_2S $ & $J_3S $ & $DS$\\ [1ex] 
 \hline
 0.060(3) & 0.110(5) & 0.020(5) & 0.012(1)\\ [1ex]   
 \hline\hline
\end{tabular}
\end{table}

In order to characterize the full magnon band structure, the energy-momentum slices were plotted in Figs.~\ref{ris:fig5}(a)--\ref{ris:fig5}(c) for three main high-symmetry directions crossing the \textit{L}-point. These are ($HHH$), $(\frac{1}{2}$~$\frac{1}{2}+H$~$\frac{1}{2}-H)$, and $(\frac{1}{2}+H$~$\frac{1}{2}+H$~$\frac{1}{2}-2H)$, which are mutually orthogonal and correspond to the $\Gamma$--\textit{L}--$\Gamma^{'}$, \textit{W}--\textit{L}--\textit{W}$^{'}$, and \textit{K}--\textit{L}--\textit{U} paths, respectively. Figure~\ref{ris:fig5}(a) represents the data on the same part of reciprocal space as previously discussed in Fig.~\ref{ris:fig3}, but with a higher resolution ($E_{\text i} = 2.68$~meV). As can be seen, two magnon modes are clearly resolved in every momentum direction. It is also evidenced that the excitations are gapped. Overall the spin waves exhibit a steeper dispersion along $\Gamma$--\textit{L} then in the two perpendicular directions, which demonstrate very similar lineshapes and stiffness. The two modes are well separated in energy for the most of \textit{W}--\textit{L} and \textit{K}--\textit{L} but approach a close energy of $\sim 0.6$~meV at the \textit{W}- and \textit{K}-points. To extract the energy of each mode at different momenta, the INS intensity profiles were fitted with Gaussian functions, as was exemplary shown in Figs.~\ref{ris:fig4}(a)--\ref{ris:fig4}(c), and plotted over intensity maps in Figs.~\ref{ris:fig5}(a)--\ref{ris:fig5}(c). The determined peak positions were used as the input for the linear spin-wave theory (LSWT) calculations using SpinW software~\cite{Toth2015}.

To reproduce the experimental magnon dispersion, we considered the model described by the following Hamiltonian:

\begin{equation}
\mathcal{H} = \sum_{\langle ij\rangle^1} J_1S_iS_j + \sum_{\langle ij\rangle^2} J_2S_iS_j + \sum_{\langle ij\rangle^3} J_3S_iS_j - \sum_i D\left( S_i^{\alpha}\right)^2,
\label{eq:eq1}
\end{equation}
where the first three terms denote the Heisenberg exchange interactions between the spins on the first ($J_1$), the second ($J_2$), and the third ($J_3$) nearest-neighbour sites, respectively. The exchange scheme is depicted in Fig.~\ref{ris:fig1}(c). The last term stands for the uniaxial single-ion anisotropy, which was included to model the spin-wave gap. The best agreement between the theory and experiment was found for the set of parameters listed in Table~\ref{tab:tab1}. The simulated spectra are shown in Figs.~\ref{ris:fig5}(d)--\ref{ris:fig5}(f) for a comparison with the experimental data. As one can see, a good agreement is achieved for both modes for all three momentum directions. Is is worth to mention, that an attempt to describe the experimental spectra with a model that includes only first two exchange interactions leads to a similarly good reproduction of the dispersion along $\Gamma$--\textit{L} but fails to agree with the spectra for \textit{W}--\textit{L} and \textit{K}--\textit{L} for any combination of $J_1$ and $J_2$. The inclusion of the $J_3$ interactions is necessary to match the observation in the vicinity of the \textit{W}- and \textit{K}-points. When the fourth nearest-neighbor Heisenberg exchange is included in the model with the optimized set of $J_1$, $J_2$, and $J_3$ parameters, no further improvement between the experimental and simulated spectra can be obtained. Therefore, it is concluded that the model with three exchange parameters is a sufficient model of the spin dynamics in GdPtBi. It is worth to mention that the exclusion of $J_3$ does not lead to a noticeable change in the optimized $J_1$ and $J_2$ parameters (the relative change is less than 5\%). In the other words, $J_1$ and $J_2$ are very weakly dependent on the strength of $J_3$.

Our LSWT calculations took into account a combinations of all four magnetic domains equally populated. The observed two modes are contributions of domains with different orientation of propagation vectors. The magnon dynamics of each domain is characterized by the dispersion that alters between the part of the momentum space that encloses the domain's propagation vector and the momentum volume corresponding to the propagation vector of any other domain. For instance, if the \textit{L}$(\frac{1}{2}\frac{1}{2}\frac{1}{2})$--$\Gamma$(111) momentum path is considered [Figs.~\ref{ris:fig5}(a),~\ref{ris:fig5}(d)], the upper mode correspond to the domain with $\mathbf{k}_3 = (\frac{1}{2}\frac{1}{2}\frac{1}{2})$, whereas the lower mode is a superposition of the dispersions of the domains with $\mathbf{k}_1$, $\mathbf{k}_2$, and $\mathbf{k}_4$. In other words, if one magnetic domain is singled out (by field-cooling procedure or strain), the magnon spectra would consist of only one branch in any point of the BZ.

Here it is interesting to note that within every domain, the low-energy spin-wave spectrum consists not only of the excitations emanating from the ordering vector (with the smaller energy gap), but also of the soft modes in three other structurally equivalent $L$ points (with a slightly larger gap). In the Heisenberg limit $D \rightarrow 0$, both gaps would vanish in LSWT, which is reminiscent of the situation recently reported in the helimagnet ZnCr$_2$Se$_4$ with a cubic spinel structure \cite{Tymoshenko2017}. There, the ordering wave vector was spontaneously chosen along one of the three equivalent cubic axes, and the Goldstone mode emanating from the ordering vector coexisted with soft magnon modes (so-called pseudo-Goldstone modes) at the two equivalent points in the structural BZ. The situation in GdPtBi is similar, yet due to the different orientation of the magnetic propagation vector along the diagonal of the BZ, there is one Goldstone and three pseudo-Goldstone modes in the BZ. This implies that the magnon density of states at low energies is dominated by the pseudo-Goldstone modes with the larger gap size ($\sim$0.25~meV), which should give rise to measurable anomalies in the temperature dependence of thermodynamic and transport properties, such as specific heat or magnon heat conduction, around 2--3~K. Such an anomaly has been indeed observed in the specific-heat data in Ref. \cite{Mueller2014} inside the magnetically ordered phase, in perfect quantitative agreement with our INS measurements. It is also very similar to the one seen in ZnCr$_2$Se$_4$ by Gu \textit{et al.} \cite{Gu2018}. Based on our present results, we can associate this anomaly with the pseudo-Goldstone magnon gap, as these modes are responsible for 3/4 of the magnon density of states in GdPtBi at low energies.

\section{discussion and conclusion}

The magnetic structure of GdPtBi was a subject of early \textit{ab initio} calculations that, as was later discovered, failed to give the correct predictions~\cite{Khmelevskyi2012,Hallouche2014}. The $\mathbf{k} = (\frac{1}{2}10)$ structure was suggested as the lowest-energy one in~\cite{Khmelevskyi2012}, whereas the $\mathbf{k} = (100)$ phase was proposed in~\cite{Hallouche2014}. In both cases, the real structure was found to have a much higher energy. In another study~\cite{Li2015}, it was concluded that the $\mathbf{k} = (\frac{1}{2}\frac{1}{2}\frac{1}{2})$ structure is the stable magnetic order in agreement with the experimental findings. However, the difference between the $\mathbf{k} = (\frac{1}{2}\frac{1}{2}\frac{1}{2})$ and $\mathbf{k} = (100)$ states was claimed to be negligible.

The exchange interactions extracted from the magnon spectra allowed us to place GdPtBi on the ($J_2/J_1$, $J_3/J_1$) phase diagram shown in Fig.~\ref{ris:fig1}(d). Its position, (1.83, 0.33), corresponds to a point deep in the $\mathbf{k} = (\frac{1}{2}\frac{1}{2}\frac{1}{2})$ phase and far from the phase boundaries of the other phases. This is in strict contrast to the results of the study~\cite{Li2015}. Despite the fact that the fcc lattice with AFM nearest-neighbor interactions is an example of a frustrated system, the next-nearest AFM coupling in GdPtBi, $J_2$, is almost two times larger than the nearest-neighbor exchange $J_1$, which drives the compound to an essentially unfrustrated condition ($f < 5$ for $J_2/J_1 > 1.3$~\cite{Ravelli2019}). This suggests that the magnetic frustration is not the origin of the large anomalous Hall angle~\cite{Suzuki2016,Shekhar2018}. The low-frustration scenario also agrees with the $\mu$SR measurements~\cite{Shekhar2018}, which did not reveal any magnetic correlation effects above $T_{\text N}$. The $J_2 > J_1$ inequality also implies a low spin reduction induced by zero-point quantum fluctuations~\cite{Datta2012,Yildirim1998}, which explains a large low-temperature magnetic moment~\cite{Shekhar2018,Mueller2014}.

Generally, a large $J_2$ is not uncommon for the $\mathbf{k} = (\frac{1}{2}\frac{1}{2}\frac{1}{2})$ fcc antiferromagnets. For instance, $J_2/J_1 \sim 3$ in EuTe and CoO, $J_2/J_1 \sim 1.8$ in $\alpha$-MnS; the compounds NiO, MnO, FeO (distorted fcc lattice) are also characterized by $J_2 > J_1$~(\cite{Datta2012} and refs. therein). The examples of the fcc AFMs that demonstrate the inverse relation $J_2 < J_1$ include double perovskites Ba$_2$CeIrO$_6$~\cite{Ravelli2019}, Ba$_2$YOsO$_6$~\cite{Kermarrec2015}, and Sr$_2$YRuO$_6$~\cite{Disseler2016}, hexahalide K$_2$IrCl$_6$~\cite{Khan2019}, and pyrite MnS$_2$~\cite{Chatterji2019}, all of them exhibit pronounced frustration effects in contrast to GdPtBi.

Seemingly, a less expected observation is that the excitations are gapped with a sizeable gap $\Delta/W \sim 0.2$, where $\Delta$ is the spin-wave gap, and $W$ is the dispersion bandwidth. It is well understood that the magnetic dipolar forces favor the magnetic moments oriented within the FM (111) planes for the fcc AFMs with the $\mathbf{k} = (\frac{1}{2}\frac{1}{2}\frac{1}{2})$ order~\cite{Rotter2003,Johnston2016}. The dipole-dipole interaction is invariant with respect to the spin rotation within the plane, which induces an effective easy-plane anisotropy. The easy-plane anisotropy splits the double-degenerate mode in the vicinity of the \textit{L}-point into two modes, one of which becomes gapped and the other one remains gapless~\cite{Batalov2016,Datta2012,Battles1969}. Thus, the additional anisotropy within the (111) plane is necessary to induce the gap in the second mode. Such an anisotropy is not typical for the half-filled 4$f$-electron systems like Gd$^{3+}$ or Eu$^{2+}$ (4$f^7$ configuration). However, recent studies demonstrated that Gd 4$f$-5$d$ hybridization leads to 4$f$-orbital anisotropy and the orbital order in GdB$_4$~\cite{Jang2016}. Besides, the rotational-symmetry breaking of the 4$f$ states was observed in EuO~\cite{Laan2008}. We argue that a similar scenario might take place in the case of GdPtBi.

To conclude, we carried out INS measurements that covered a large part of the 4D energy-momentum space. The collected magnon spectra allowed us to identify a gapped dispersive mode and resolve its dispersion across the entire BZ. The observed spectra were simulated within the LSWT approach, which enabled us to construct an effective Heisenberg Hamiltonian that describes the magnetic dynamics in GdPtBi. The determined exchange interactions agree with the previously reported ground state and indicate the absence of strong magnetic frustration in the material.

\section*{Acknowledgments}

A.S.S. thanks S. E. Nikitin for stimulating discussions. This project was funded in part by the German Research Foundation (DFG) under Grant No. IN 209/7-1, via the project C03 of the Collaborative Research Center SFB 1143 (project-id 247310070) at the TU Dresden and the W\"{u}rzburg-Dresden Cluster of Excellence on Complexity and Topology in Quantum Matter -- \textit{ct.qmat} (EXC 2147, project-id 39085490). A.S.S. acknowledges support from the International Max Planck Research School for Chemistry and Physics of Quantum Materials (IMPRS-CPQM).

\end{document}